# Design of Fiber-Longitudinal Optical Power Monitor

Takeo Sasai, *Member, IEEE, Member, Optica*, Sze Yun Set, *Member, IEEE, Fellow, Optica,* Shinji Yamashita, *Member, IEEE, Fellow, Optica*

*Abstract*—**This paper presents analytical results on the accuracy of fiber-longitudinal optical power monitoring (LPM) at arbitrary positions. To quantify the accuracy, the position-wise variance and power-profile SNR of LPM are defined and analyzed, yielding formulas for these metrics. Using these metrics, we show that various designs and performance predictions of LPM for a given link and estimation conditions are possible in a unified manner. Specifically, the required SNR to detect a given loss event is first presented. Based on this, the design parameters of LPM, such as the sample size and optical power required to detect the loss, are explicitly determined. The performance such as the detectable limit of loss events at individual positions and maximum dynamic range are also specified. These results can be used as a basis for establishing a design principle of LPM.**

## I. INTRODUCTION

IN fiber-optic communications, monitoring of transmission link parameters is essential to maximize transmission capacity with less margin and ensure stable system operation [1]. The optical power is one of the fundamental parameters to monitor as it dominantly determines the generalized signal-to-noise ratio (GSNR) and thus the capacity and reach [2]. Monitoring of optical power allows for GSNR estimation [3][4], optimal selection of transmission modes [5], fault localization, launch power optimization, and proactive maintenance of networks.

The fiber-longitudinal power monitor (LPM), intensively studied in recent years, enables a distributed measurement of optical power (precisely, the product of nonlinear constant and optical power) over the entire link solely by processing signals received at a coherent receiver [6][7][8][9][10][11][12][13][14]. As a result, fiber losses, amplifier gains, and loss anomalies can be inspected in an end-to-end manner. Since LPM does not rely on dedicated hardware devices, it allows for ubiquitous network monitoring without being influenced by network configurations, equipment vendors, and domain structures [5]. By extending algorithms of LPM, distributed measurements of numerous parameters have been demonstrated such as chromatic dispersion (CD) map [7], fiber types [15], responses of optical filters [7][16], polarization dependent losses [17][18][19], spectral tomography over C [7][20], C+L [21], and S+C+L bands [22], 4D optical link tomography [23], multi-path interference [24], and differential group delay [25]. The performance of LPM has been drastically enhanced as can be seen in demonstrations including a precise estimation closely matching optical time domain reflectometer (OTDR) [13], LPM over 10,000 km [9], and field demonstrations under the system optimal launch power with a full C-band wavelength division multiplexing (WDM) using commercial transponders [23]. The theoretical background has been gradually developed, such as performance limit, spatial resolution, and the difference between LPM methods [8][26][27].

One practically important yet not fully understood aspect of LPM is its accuracy in the presence of noise. Since LPM relies on fiber Kerr nonlinearity to estimate optical power, the accuracy depends on optical power and, consequently, on positions. A previous work has analyzed the performance of LPM methods in noise-less conditions, but the behavior under noise has been limited to qualitative discussions [8]. In [27], the performance of multiple LPM methods in the presence of noise has been quantitatively compared in terms of the overall accuracy averaged over positions. Despite these efforts, the following aspects of LPM remain unclear: For given parameters including signal power $P(z)$ and noise power $\sigma^2$,

- how is the accuracy of LPM at arbitrary positions $z$ expressed?
- what is the detectable limit of loss anomalies at arbitrary positions $z$?
- how many signal samples are required to achieve a desired accuracy?

This paper aims to answer these questions. Specifically, to quantify the accuracy at individual positions, the position-wise variance and SNR of the estimated power at individual positions in the presence of noise are defined and analyzed in Section II, leading to formulas (21) and (22) for these metrics. Through this analysis, using Szegö's and Tyrtyshnikov's theorems, these metrics are found to be effectively explained by the discrete time Fourier transform (DTFT) of the spatial correlation function (SCF) that has been used to explain the spatial resolution of LPM in a previous work [8]. This finding leads to the derivation of an upper bound of the variance (29) in the case of Gaussian spectral signals. Section III describes how the variance and SNR defined and analyzed in Section II can be used to design the parameters of LPM and predict its performance. To this end, we first present the required SNR to detect a given loss event. This relation is then used to select LPM parameters such as the sample size and optical power required to detect the loss under a given estimation conditions. Similarly, the performance such as the detectable limit and dynamic range are also specified. These results help to design LPM and define its specifications. Section IV discusses the effects unaccounted for in our analysis, including the SNR limit of power profiles, modulation format, and regularization. Section V concludes the paper.

## II. VARIANCE AND SNR OF LONGITUDINAL POWER ESTIMATION

### A. Model

Consider the normalized nonlinear Schrödinger equation:

$$\frac{\partial A}{\partial z} = \left(j\frac{\beta_2(z)}{2}\frac{\partial^2}{\partial t^2} + \frac{\beta_3(z)}{2}\frac{\partial^3}{\partial t^3}\right)A - j\gamma'(z)|A|^2 A \quad (1)$$

$$\gamma'(z) \equiv \gamma(z)P(z)$$
$$= \gamma(z)P(0)\exp\left(-\int_0^z \alpha(z')\,dz'\right) \quad (2)$$

where $A \equiv A(z,t)$, $\alpha(z)$, $\beta_2(z)$, $\beta_3(z)$, $\gamma(z)$, and $P(z)$ are the normalized optical signals at position $z$ and time $t$, fiber loss, second/third group velocity dispersion, nonlinear constant, and optical signal power at $z$, respectively. A single polarization transmission is assumed for simplicity. In Eq. (1), the power of signal $A(z,t)$ is normalized to unity and in turn the actual power in fibers is expressed by $P(z)$ [7]. The goal of LPM is to estimate $\gamma'(z) = \gamma(z)P(z)$ from transmitted and received signals. Note that this does not mean that the transmitted signal needs to be known a priori as it can be recovered in the receiver.

The analysis presented here follows a model used in [8][13], with a noise term incorporated. Let us consider the discretized regular perturbation model of (1) [29][30] with lumped noise at a receiver. A $N$-sample received signal vector $A(L) = [A(L,0), A(L,T), \dots, A(L,(N-1)T)]^T$ at the link end $z = L$ with a sampling rate $1/T$ is then expressed as follows:

$$A(L) \simeq A_0(L) + A_1(L) + \nu \quad (3)$$

where $A_0$ is the linear term and $A_1$ is the first order perturbation term with

$$A_0(L) = D_{0L}A(0), \quad (4)$$

$$A_1(L) = G\gamma'$$
$$= [g_0, \dots, g_m, \dots, g_{M-1}]\gamma', \quad (5)$$

$$g_m \equiv -j\Delta z D_{z_m L}\widetilde{N}[D_{0z_m}A(0)], \quad (6)$$

$z_m$ is the $m$-th measurement position of LPM with total measurement points $M$, $D_{z_m z_l}$ is a matrix representing a CD

from $z_m$ to $z_l$, $\Delta z = z_{m+1} - z_m$ is the spatial granularity, $\widetilde{N} = (|\cdot|^2 - 2P)(\cdot)$ is a nonlinear operator, $P(=1)$ is the power of $A$. For the term $-2P$, see [31][8]. Our estimation target is $\gamma' = [\gamma'_0, \gamma'_1, \dots, \gamma'_{M-1}]^T$, where $\gamma'_m \equiv \gamma'(z_m)$. $\nu$ is the additive noise including link noise and deviations from the model and is assumed to follow a circularly symmetric complex Gaussian process $\nu \sim \mathcal{CN}(0, \Sigma_\nu)$ with $\Sigma_\nu$ being a covariance matrix. By rewriting (3), a linear model with respect to $\gamma'$ is obtained as

$$y \simeq G\gamma' + \nu, \quad (7)$$

where $y \equiv A(L) - A_0(L)$ is an observation vector obtained from received and transmitted waveforms using (4).

### B. Estimator of Longitudinal Optical Power

In literature, several estimators have been proposed. The original correlation method (CM) [6], modified CM [8][24], gradient descent optimization of split-step method [7], linear least squares [13] and its modifications [10][14][32]. In this work, we first adopt the penalized least squares [14], proposed as a generalization of CM and linear least squares. Considering $\gamma'$ is a real-valued vector, a penalized least squares estimator of $\gamma'$ in (7) is

$$\widehat{\gamma'} = \mathrm{Re}[G^\dagger G + \lambda R]^{-1}\mathrm{Re}[G^\dagger y], \quad (8)$$

where $\lambda$ is a regularization parameter and $R$ is a M × M matrix. This estimator introduces a further generalization of the original one [14], where $R = I$ was used, with $I$ being the identity matrix. In many applications, $R$ is often a positive semidefinite matrix, which corresponds to adding a penalizing functional with a quadratic form to the squared error cost function. While we start from (8) and present its variance in a general form, we will then assume $\lambda = 0$, i.e., linear least squares [13], to promote analysis.

### C. Position-wise Variance and SNR

To quantify the accuracy of LPM at arbitrary positions, we aim to analyze the variance of the estimates $\widehat{\gamma'}$ and then define the SNR. (8) can be divided into two terms using (7) as

$$\widehat{\gamma'} = \widehat{\gamma'}_{ideal} + \varepsilon \quad (9)$$

where $\widehat{\gamma'}_{ideal}$ is the ideal result in the absence of noise ($\nu = 0$) and $\varepsilon$ is the noise term in LPM. For the estimator (8),

$$\varepsilon = \mathrm{Re}[G^\dagger G + \lambda R]^{-1}\mathrm{Re}[G^\dagger \nu]. \quad (10)$$

Assuming $\Sigma_\nu = \sigma^2 I$ with $\sigma^2$ being a scalar noise power, $\varepsilon$ again follows the zero-mean Gaussian process $\varepsilon \sim \mathcal{N}(0, \Sigma_\varepsilon)$, where the covariance is (see Appendix A):

$$\Sigma_\varepsilon = \frac{\sigma^2}{2}\mathrm{Re}[G^\dagger G + \lambda R]^{-1}\mathrm{Re}[G^\dagger G]\mathrm{Re}[G^\dagger G + \lambda R]^{-1} \quad (11)$$

The mean and covariance of $\widehat{\gamma'}$ are therefore

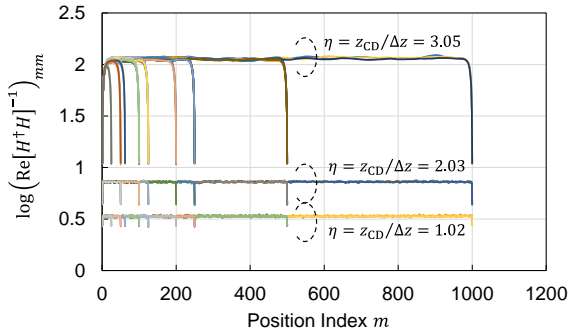

Fig. 1. Diagonal entries of $\text{Re}[H^\dagger H]^{-1}$ for various combinations of symbol rate, total distance $L$, $\Delta z$, and $\beta_2$. Rectangular spectral signals are used. The diagonal values are effectively explained by $\eta = z_{CD}/\Delta z = \frac{1}{4|\beta_2|\text{BW}^2\Delta z}$.

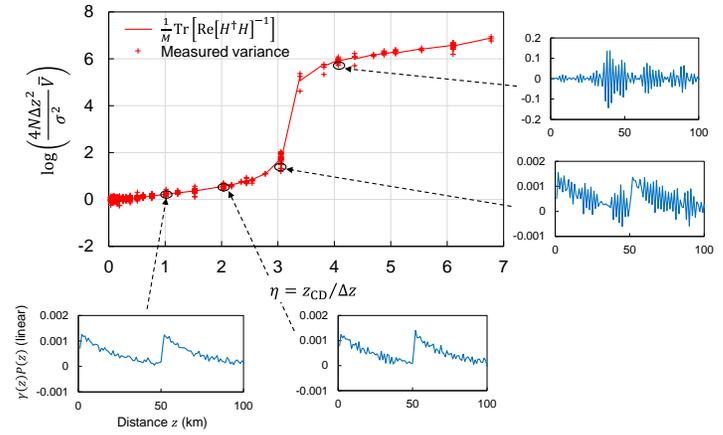

Fig. 2. Derived variance $\frac{1}{M}\text{Tr}(\text{Re}[H^\dagger H]^{-1})$ and measured (normalized) variance $\frac{4N\Delta z^2}{\sigma^2}\bar{V}$ (average over positions) of LPM for various combinations of symbol rate BW, signal samples $N$, total distance $L$, $\Delta z$, and $\beta_2$. Rectangular spectral signals are used. M: matrix size.

$$\widehat{\gamma'} \sim \mathcal{N}(\widehat{\gamma'}_{ideal}, \Sigma_\epsilon) \qquad (12)$$

where, using (9), the mean $\mathbb{E}[\widehat{\gamma'}] = \widehat{\gamma'}_{ideal}$ and the covariance $\mathbb{E}[([\widehat{\gamma'}] - \mathbb{E}[\widehat{\gamma'}])([\widehat{\gamma'}] - \mathbb{E}[\widehat{\gamma'}])^\dagger]$ are used. In what follows, we set $\lambda = 0$, resulting in $\Sigma_\epsilon = \frac{\sigma^2}{2}\text{Re}[G^\dagger G]^{-1}$. Our interest is the variance of each $\widehat{\gamma'_m}$ at a position $z_m$. They are on the diagonal of the covariance matrix as

$$V[\widehat{\gamma'_m}] = \frac{\sigma^2}{2}(\text{Re}[G^\dagger G]^{-1})_{mm}$$
$$= \frac{\sigma^2}{4N\Delta z^2}(\text{Re}[H^\dagger H]^{-1})_{mm} \qquad (13)$$

Since $G^\dagger G$ explicitly depends on sample size $N$ and squared spatial granularity $\Delta z^2$ due to (6), a normalized matrix $H = G/(\sqrt{2N}\,\Delta z)$ is introduced in (13) so that the maximum entry of $\text{Re}[H^\dagger H]^{-1}$ becomes unity. By doing so, as shown in subsequent sections, $H^\dagger H$ becomes a function primarily of $\eta = z_{CD}/\Delta z$ (with a slight dependence on matrix size $M$), where $z_{CD} = \frac{1}{4\beta_2\text{BW}^2}$ is a characteristic length of CD for rectangular signals (i.e., Nyquist limit) and BW is the signal bandwidth. This will ease the subsequent analysis.

In this paper, using (13), the position-wise SNR of estimated power profiles are defined as follows:

$$\text{SNR}_{pp}(z_m) \equiv \frac{\gamma'^2_m}{V[\widehat{\gamma'_m}]}$$
$$= \frac{4N\Delta z^2\gamma^2(z_m)P^2(z_m)}{\sigma^2(\text{Re}[H^\dagger H]^{-1})_{mm}} \qquad (14)$$

By adopting this definition, the detectable limit of a loss anomaly at arbitrary positions can be directly expressed as described in Section III.

### D. Variance is almost uniform over positions

To promote the analysis of the variance and SNR, the following assumptions are made:

1. $(\text{Re}[H^\dagger H]^{-1})_{mm}$ is uniform across positions $z_m$.
2. The transmitted signal follows a stationary and circularly symmetric complex Gaussian process.

By the assumption 1, individual diagonal entries in $\text{Re}[H^\dagger H]^{-1}$ can be represented by an average of all diagonal entries, i.e., the trace divided by a matrix size $M$. (13) is then rewritten

$$V[\widehat{\gamma'_m}] = \frac{\sigma^2}{4N\Delta z^2}\frac{1}{M}\text{Tr}(\text{Re}[H^\dagger H]^{-1}).$$
$$= \frac{\sigma^2}{4N\Delta z^2}\frac{1}{M}\sum_{m=0}^{M-1}\frac{1}{\phi_{r,m}} \qquad (15)$$

where $\phi_{r,m}$ are the eigenvalues of $\text{Re}[H^\dagger H]$, where the subscript $r$ implies $\text{Re}[\cdot]$. In (15), we used (i) the trace of a matrix is equivalent to a sum of eigenvalues, and (ii) eigenvalues of an inverse matrix are reciprocals of eigenvalues of the original matrix. Note that $\phi_{r,m} > 0$ since $\text{Re}[H^\dagger H]$ is a positive definite matrix as long as CD is monotonically accumulated with distance (i.e., dispersion-uncompensated link) [8][13].

The uniformness of the diagonals of $(\text{Re}[H^\dagger H]^{-1})_{mm}$ is shown by numerical results in Fig. 1. The diagonals are shown for various combinations of related parameters including BW, total distance $L$, $\Delta z$, and $\beta_2$. Note that rectangular spectral signals (Nyquist limit) are used and thus BW is equal to symbol rate. $H$ is calculated by $H = G/(\sqrt{2N}\,\Delta z)$, where $G$ is based on (5) and (6). The sweep ranges of parameters are described in Appendix B. There are three notable points: (i) The diagonal values are almost determined by $\eta$, irrespective of the total distance $L$ or matrix size. (ii) For small $\eta$, the diagonals are almost flat over positions except for edges, while increasing $\eta$ results in degraded flatness (see $\eta = 3.05$). (iii) However, increasing $\eta$ simultaneously results in higher diagonal values, implying greater variances. For instance, raising $\eta$ from 2.03 to 3.05 boosts the variance nearly tenfold. This suggests that operating LPM around $\eta = 3.05$ is not advisable due to the significant noise enhancement, as will be observed later in Fig. 2. It is therefore reasonable to assume that LPM operates at small $\eta$, where the variance remains suppressed and uniform across positions.

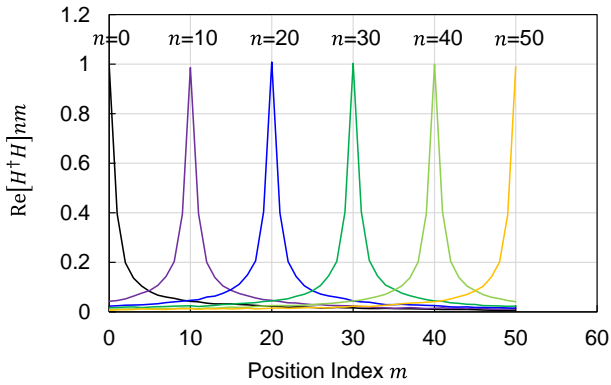

Fig. 3. $n$-th rows of $\mathrm{Re}[H^{\dagger}H]$ for Gaussian signal format. Each row is a shifted spatial correlation function (SCF). $\beta_2$=-21.0 ps²/km, BW = 128 GHz, $\Delta z = 1.0$ km, $L = 50$ km, and rectangular spectral signals are used.

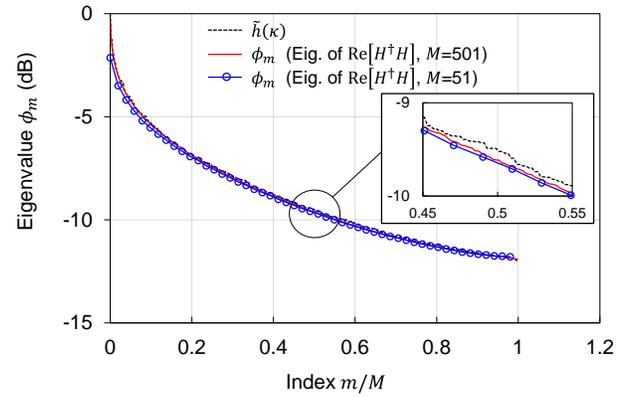

Fig. 4. Comparison of eigenvalues of $\mathrm{Re}[H^{\dagger}H]$ and Fourier transform of spatial correlation function (SCF), sorted in nonincreasing order. $m$: index of eigenvalues, $M$: matrix size, $\beta_2$=-21.0 ps²/km, $\Delta z = 1.0$ km, BW = 128 GHz. Rectangular spectral signals are used.

Based on these observations, the variances at individual positions are represented by the trace of $\mathrm{Re}[H^{\dagger}H]^{-1}$ in (15). Fig. 2 plots the derived $\frac{1}{M}\mathrm{Tr}[\mathrm{Re}[H^{\dagger}H]^{-1}]$ and normalized variance $\frac{4N\Delta z^2}{\sigma^2}\bar{V}$ measured from estimated power profiles in simulations for various parameters. The measured variance is a statistical measure (average) over positions and, due to the assumption 1, it can also be regarded as the variance at each position. The detail of the simulations and the sweeping range of the parameters are described in Appendix B. Four power profiles are also shown in insets, corresponding to $\eta = 1.02$, 2.03, 3.05, and 4.07. $\eta$ is used for the horizontal axis as it is an effective metric to describe the evolution of the variance, as discussed above. The trace metric agrees well with the observed variance for a wide range of $\eta$. At around $\eta = 3.05$, the variance rapidly grows, leading to strong noise enhancement and instable estimation of power profiles, which was also observed in Fig. 1. This observation is discussed in [13] using the condition number of $G$ and is attributed to the increased ill-posedness of the least squares estimation due to reduced CD effect in $\Delta z$. For large $\eta$ ($\eta = z_{\mathrm{CD}}/\Delta z$), the CD effect in $\Delta z$ decreases, and the signal waveforms at two positions with $\Delta z$ apart do not alter significantly. As a result, the Kerr nonlinear waveforms excited at these positions are indistinguishable, leading to an increased uncertainty of the estimates and noisy power profiles.

The limitation of this uniform variance assumption will be discussed in Section IV.B.

### E. Approximation using Szegö's theorem

Under the assumption 2, it has been shown [8][13] that $H^{\dagger}H$ (and thus also $\mathrm{Re}[H^{\dagger}H]$) becomes a Toeplitz matrix (i.e., a linear convolution operator)

$$H^{\dagger}H = \begin{bmatrix} h_0 & h_{-1} & h_{-2} & \cdots & h_{-(M-1)} \\ h_1 & h_0 & h_{-1} & & \\ h_2 & h_1 & h_0 & & \vdots \\ \vdots & & & \ddots & h_{-1} \\ h_{M-1} & \cdots & & h_1 & h_0 \end{bmatrix} \quad (16)$$

where $h_m$ is the so-called (normalized) spatial correlation function (SCF) or spatial response function [8][26]. Fig. 3 shows an example of the rows of $\mathrm{Re}[H^{\dagger}H]$ for a Gaussian signal format, illustrating that each row is a shifted SCF. The SCF essentially determines the performances of the LPM, including spatial resolution as discussed in [8] and SNR as in this paper. Based on this Toeplitz assumption and (15), our problem reduces to evaluating the eigenvalues $\phi_m$ of the Hermitian Toeplitz matrix $\mathrm{Re}[H^{\dagger}H]$. The impact of other modulation formats such as QPSK will be discussed in Section IV.B.

According to Szegö' theorem [33] and its extensions [34][35], the eigenvalues of the sequences of Hermitian Toeplitz matrices can be related to Fourier series:

$$h_m = \int_{-\pi}^{\pi} \tilde{h}(\kappa) e^{-jm\kappa} d\kappa, \qquad m \in \mathbb{Z} \quad (17)$$

$$\tilde{h}(\kappa) = \sum_{m=-\infty}^{\infty} h_m e^{jm\kappa}, \qquad \kappa \in [-\pi, \pi] \quad (18)$$

where $\kappa$ is a normalized wavenumber. $\tilde{h}$ is called a symbol or a generating function of a Toeplitz and can be understood as the discrete time Fourier transform (DTFT) of $h_m$ (or more precisely, the discrete spatial Fourier transform, though we will continue referring to it as DTFT hereafter). Specifically, Szegö's theorem states that, for $\tilde{h}(\kappa) \in L^{\infty}([-\pi, \pi])$,

$$\lim_{M \to \infty} \frac{1}{M} \sum_{m=0}^{M-1} f(\phi_m) = \frac{1}{2\pi} \int_{-\pi}^{\pi} f(\tilde{h}(\kappa)) d\kappa \quad (19)$$

where $f$ is any function continuous on the range of $\tilde{h}$, and $\phi_m$ are eigenvalues of a Hermitian Toeplitz matrix with a matrix size $M$. This theorem cannot be directly applied to our problem because $\tilde{h}$ is not necessarily in $L^\infty$, as we will see later through an example. Tyrtyshnikov [34], however, extended this theorem for $\tilde{h}(\kappa) \in \mathbb{R}$ and $\tilde{h}(\kappa) \in L^2([-\pi, \pi])$ using $f(x)$ with a compact support. Note also that, in our case, we should consider (19) for the real part of the SCF, i.e., the eigenvalues of Re$[H^\dagger H]$ and the DTFT of Re$[h_m]$, which are denoted by $\phi_{r,m}$ and $\tilde{h}_r(\kappa)$, respectively.

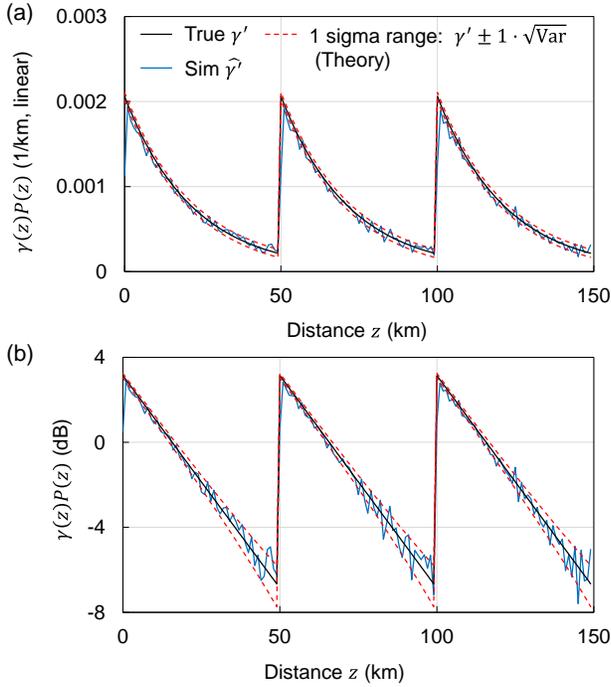

Fig. 5. Estimated longitudinal power in presence of noise (blue solid) and analytical prediction of 1 sigma range $\gamma' \pm 1 \cdot \sqrt{\text{Var}}$ (red dashed) on (a) linear and (b) logarithmic scale. $\beta_2 = -21.0$ ps$^2$/km, $\gamma = 1.0$ W$^{-1}$km$^{-1}$, $\Delta z = 1.0$ km, BW = 128 GHz, $L = 150$ km, launch power $P(0) = 2.0$ dBm, received SNR = 17.0 dB, $N = 3.1$e6, and rectangular spectral signals are used.

Szegő's and Tyrtyshnikov's theorems above describe an averaging behavior of eigenvalues and do not concern individual eigenvalues. Nevertheless, in our case, a good agreement between individual eigenvalues and the DTFT of the SCF can be observed. Fig. 4 shows the eigenvalue distributions of a Hermitian Toeplitz Re$[H^\dagger H]$ for matrix sizes of $M = 51$ and 501. We used the DFT of Re$[h_m]$ for $M = 501$ as a discretized version of the DTFT $\tilde{h}_r(\kappa)$. While (19) requires $M \to \infty$, $\tilde{h}(\kappa)$ serves as a good approximation of the eigenvalues of $H^\dagger H$ even for a small matrix size $M = 51$.

Based on these theorems and observations, we approximate (15) using (19) for the real part of the SCF $\tilde{h}_r(\kappa) = \mathcal{F}[\text{Re}[h_m]]$. By choosing $f(x) = \frac{1}{x}$ in (19), we obtain

$$\lim_{M \to \infty} \frac{1}{M} \sum_{m=0}^{M-1} \frac{1}{\phi_{r,m}} = \frac{1}{2\pi} \int_{-\pi}^{\pi} \frac{1}{\tilde{h}_r(\kappa)} d\kappa \qquad (20)$$

Then the variance (13) and SNR (14) are expressed as

$$V[\widetilde{\gamma_m'}] = \frac{\sigma^2}{8\pi N \Delta z^2} \int_{-\pi}^{\pi} \frac{1}{\tilde{h}_r(\kappa)} d\kappa \qquad (21)$$

$$\text{SNR}_{\text{pp}}(z_m) = \frac{8\pi N \Delta z^2 \gamma^2(z_m) P^2(z_m)}{\sigma^2 \int_{-\pi}^{\pi} \frac{1}{\tilde{h}_r(\kappa)} d\kappa} \qquad (22)$$

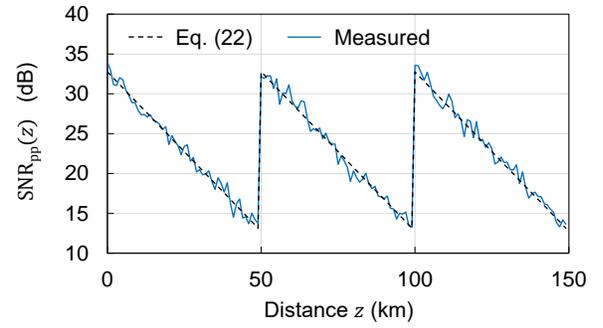

Fig. 6. Position-wise power-profile SNR. 50 power profiles are used to calculate the measured SNR. The conditions are the same as in Fig. 5.

These equations indicate that, the accuracy of LPM is essentially determined by the Fourier transform of the SCF, as well as obvious parameters such as the noise power, signal power, nonlinear constant, and sample size. Since the SCF is governed by the CD effect induced over a step size $\Delta z$ [8], the variance and SNR of power profiles also depends on CD parameters $\beta_2$, signal bandwidth BW, and spatial step size $\Delta z$. Indeed, $\int_{-\pi}^{\pi} 1/\tilde{h}_r(\kappa) d\kappa$ is a function of $\eta = z_{CD}/\Delta z$, as partly discussed above and shown soon later. Using these equations (21) and (22), it is possible to design LPM, selecting appropriate LPM parameters for a given link parameters (see Section III).

Fig. 5 shows an example of results of LPM, with the prediction of 1 sigma range $\gamma'_{ideal} \pm 1 \cdot \sqrt{V(\gamma'_m)}$ using (21). Note that the noise in LPM follows a Gaussian process as shown in (12). Simulation conditions are based on the reference system configuration described in Appendix B. Since the variances are constant across positions on a linear scale, power profiles are overwhelmed by noise in low signal power regions. This is more clearly observed on a logarithmic scale. To confirm that observations statistically match the predicted SNR (22), 50 power profiles are prepared. Their ensemble SNR is shown by a solid line in Fig. 6, showing good agreement with (22) (dashed line). The numerous observations and discussions in literature that the accuracy of LPM is degraded in lower power region due to insufficient Kerr nonlinearity can now be quantitatively understood by (21) and (22).

Now we explore a special case.

### F. Example: Gaussian Spectral Signal

In [8], it has been shown that, if the signal has a Gaussian spectrum $\tilde{p}_A(\omega) \propto \exp\left(-\frac{\omega^2}{2\sigma_0^2}\right)$, the expectation of the SCF $h_m$ is then a square root of a complex-valued Lorentzian function (see Fig. 7(a)):

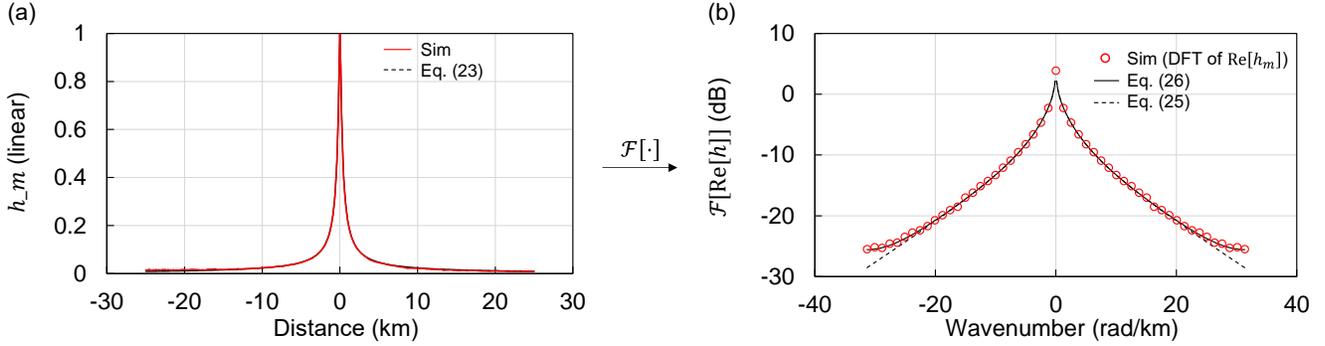

Fig. 7. Example of (a) Spatial correlation function (SCF) and (b) its Fourier transform. Matrix size: 501, $\Delta z = 0.1$ km, $\beta_2 = -21.0 \text{ps}^2/\text{km}$, $\sigma_\omega/2\pi = 54.4$ GHz, Gaussian spectral signals are used. In (b), simulation plots are decimated for visibility.

$$h_m = \frac{1}{\sqrt{1 + 2j\left(\frac{m\Delta z}{z_{CD}}\right) + 3\left(\frac{m\Delta z}{z_{CD}}\right)^2}}$$
$$= \frac{1}{\sqrt{1 + 2j(m/\eta) + 3(m/\eta)^2}} \qquad (23)$$

where, in a Gaussian spectral case, $z_{CD} = \frac{1}{|\beta_2|\sigma_\omega^2}$. Note that, $h_m$ is square summable, $\sum_{m=-\infty}^{\infty} |h_m|^2 < \infty$. To further proceed, we first consider a continuous version of (23): $h(z) = \frac{1}{\sqrt{1+2j(z/z_{CD})+3(z/z_{CD})^2}}$ ($z \in \mathbb{R}$). Since the Fourier transform of the square root of a Lorentzian is $\mathcal{F}\left[\frac{1}{\sqrt{1+t^2}}\right] = 2K_0(|\omega|)$ [36], the Fourier transform of $h(z)$ is then (see Appendix C):

$$\bar{h}(k) = \mathcal{F}[h(z)]$$
$$= \frac{2}{\sqrt{3}} z_{CD} K_0\left(\frac{2}{3} z_{CD}|k|\right) \exp\left(-\frac{1}{3} z_{CD}k\right) \qquad (24)$$

where $k \in \mathbb{R}$ is an angular wavenumber in rad/m and $K_0(x)$ is the modified Bessel function of the second kind. Then the real-part case is also obtained as (see Appendix C):

$$\bar{h}_r(k) = \mathcal{F}[\text{Re}[h(z)]]$$
$$= \frac{2}{\sqrt{3}} z_{CD} K_0\left(\frac{2}{3} z_{CD}|k|\right) \cosh\left(\frac{1}{3} z_{CD}k\right) \qquad (25)$$

Considering an aliasing effect, the DTFT of the original discretized SCF $\text{Re}[h_m]$ is therefore expressed as

$$\tilde{h}_r(\kappa) = \frac{1}{\Delta z} \sum_{l=-\infty}^{\infty} \bar{h}_r\left(\frac{\kappa}{\Delta z} + \frac{2\pi l}{\Delta z}\right), \qquad \kappa \in [-\pi, \pi] \qquad (26)$$

Note that, according to (25) and (26), $\tilde{h}_r(\kappa)$ goes to infinity as $\kappa \to 0$, and thus $\tilde{h}_r(\kappa) \notin L^\infty([-\pi, \pi])$. However, $\tilde{h}_r(\kappa) \in L^2([-\pi, \pi])$ due to the square summability of $h_m$ and the Parseval's identity.

Fig. 7(b) shows examples of (25) and (26), and the DFT of the numerical $h_m$ constructed from signals using (5) and (6). When plotting $\tilde{h}_r(\kappa)$, the axis is transformed to $k = \kappa/\Delta z$ in rad/km. $\tilde{h}_r(\kappa)$ is always larger than $\bar{h}_r(k)$ at the edge (Nyquist 'wavenumber') due to the aliasing. Since we will take the inverse of $\tilde{h}_r(\kappa)$ to apply (20), our interest is a region where $\tilde{h}_r(\kappa)$ is small, i.e., $\kappa$ is large. To further proceed, we consider the following two approximations

- Consider only the zero-th term ($l = 0$) in (26), ignoring the aliasing.
- Apply an asymptotic expansion of (25):

$$\bar{h}_r(k) \sim \sqrt{\frac{\pi}{12}} z_{CD} \frac{\exp\left(-\frac{1}{3} z_{CD}|k|\right)}{\sqrt{\frac{1}{3} z_{CD}|k|}}, \qquad z_{CD}|k| \to \infty \qquad (27)$$

where we used an asymptotic expansion of the Bessel function $K_0(x) \sim \sqrt{\frac{\pi}{2x}} e^{-x}$ ($x \to \infty$) [36] and ignored a $e^{-3x}$ term in the resulting $e^{-x} + e^{-3x}$ of $K_0(2x)\cosh(x)$. For both approximations, the error is increased as $\eta \to 0$, i.e., increased $\Delta z$ or CD $|\beta_2|\sigma_\omega^2$. For the first approximation, this is due to the enhanced aliasing effect. For the second approximation, this is because, when $\eta$ is decreased, $z_{CD}|k|(= \eta|\kappa|)$ is also reduced, making the asymptotic expansion an inadequate approximation. This will be observed in Fig. 8.

Now we evaluate the integral in (20). Since the Fourier transform of a real-valued even function is a real-valued even function, we only consider $\kappa > 0$. Then (see Appendix D)

$$\lim_{M \to \infty} \frac{1}{M} \sum_{m=0}^{M-1} \frac{1}{\phi_{r,m}} = \frac{1}{\pi} \int_0^\pi \frac{1}{\tilde{h}_r(\kappa)} d\kappa$$
$$< 2\left(\frac{3}{\pi}\right)^{\frac{3}{2}} \frac{\Delta z^2}{z_{CD}^2} \left| \Gamma\left(\frac{3}{2}, -\frac{\pi z_{CD}}{3\Delta z}\right) \right| \qquad (28)$$
$$= 2\left(\frac{3}{\pi}\right)^{\frac{3}{2}} \frac{1}{\eta^2} \left| \Gamma\left(\frac{3}{2}, -\frac{\pi}{3}\eta\right) \right|$$

where $\Gamma(a,x)$ is the lower incomplete gamma function. (28) shows that $\int_{-\pi}^{\pi} 1/\tilde{h}_r(\kappa) d\kappa$ is a function of $\eta$, although this is already clear from (23), where the original $h_m$ depends only on $\eta$. Finally, from (13) and (14), the variance and SNR of estimated longitudinal power satisfy:

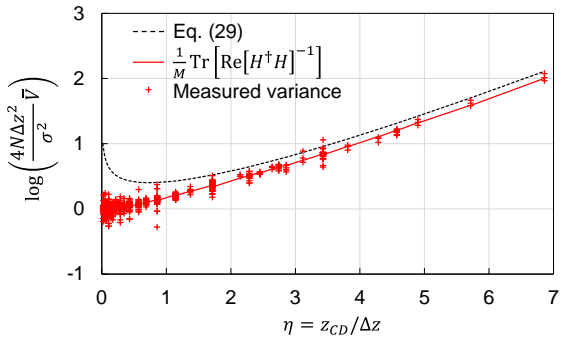

Fig. 8. Derived and measured (normalized) variance $\frac{4N\Delta z^2}{\sigma^2} \bar{V}$ (average over positions) of LPM for various combinations of symbol rate BW, signal samples $N$, total distance $L$, $\Delta z$, and $\beta_2$. Gaussian spectral signals are used.

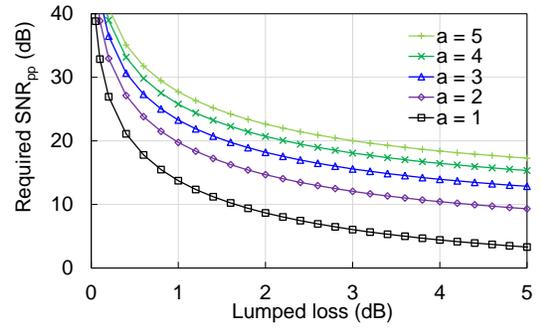

Fig. 9. Required $\mathrm{SNR_{pp}}$ to detect a given lumped loss that satisfies $\gamma' - a\sqrt{V} > \mathrm{loss} \cdot \gamma'$.

$$\mathrm{V}[\widehat{\gamma_m}] < \frac{1}{2}\left(\frac{3}{\pi}\right)^{\frac{3}{2}} \frac{\sigma^2}{N z_{CD}^2}\left|\Gamma\left(\frac{3}{2}, -\frac{\pi z_{CD}}{3\Delta z}\right)\right| \qquad (29)$$

$$\mathrm{SNR_{pp}}(z_m) > 2\left(\frac{\pi}{3}\right)^{\frac{3}{2}} \frac{N z_{CD}^2 \gamma^2(z_m) P^2(z_m)}{\sigma^2 \left|\Gamma\left(\frac{3}{2}, -\frac{\pi z_{CD}}{3\Delta z}\right)\right|} \qquad (30)$$

Fig. 8 shows the derived and measured normalized variance $\frac{4N\Delta z^2}{\sigma^2} \bar{V}$ obtained by LPM simulations for Gaussian spectral signals, similarly to Fig. 2 for rectangular spectral signals. Again, $\bar{V}$ is a statistic over positions, assuming the position-wise variance is uniform across positions. A theoretical line based on the trace metric agrees well with the numerical results as observed in the case of rectangular spectral signals (Fig. 2). The dashed line is an upper bound (29), and asymptotically approaches to the observation as $\eta = z_{CD}/\Delta z \to \infty$, while a large deviation observed for $\eta \to 0$. This is primarily due to ignoring the aliasing effect that becomes significant when the spatial sampling rate $1/\Delta z$ is low for a given $z_{CD}$. Ignoring the aliasing effect results in an underestimation of $\tilde{h}_r(\kappa)$ (see Fig. 7(b)), which leads to an overestimation of the integral of its reciprocal.

## III. DESIGN OF LONGITUDINAL POWER ESTIMATOR

### A. Required SNR to Detect a Loss

Using the SNR defined in Section II, the detectable limit of a loss event can be directly expressed. Let $\gamma'$ be a power at a certain position and $\mathrm{loss} \cdot \gamma'$ be a power after a lumped loss, where $0 < \mathrm{loss} < 1$ on a linear scale. In this paper, the detectable loss is defined as loss satisfying

$$\gamma' - a\sqrt{V} > \mathrm{loss} \cdot \gamma' \qquad (31)$$

where $a$ is a scalar design parameter that determines the confidence level of detection. A physical meaning is that, in order to detect a loss event, power difference due to a loss event $\gamma' - \mathrm{loss} \cdot \gamma'$ should at least be greater than $a$-sigma range. Since noise in LPM follows a Gaussian process due to (12), the left-hand side of (31) is a lower $a$-sigma of a Gaussian distribution. Transforming (31) yields the required SNR to detect a given loss event:

$$\mathrm{SNR_{pp}}(z_m) > \left(\frac{a}{1 - \mathrm{loss}}\right)^2 \qquad (32)$$

where the definition of $\mathrm{SNR_{pp}}$ (14) was used. Fig. 9 shows (32) for various $a$. As the detection capability criteria become more demanding, the required $\mathrm{SNR_{pp}}$ correspondingly grows. For instance, if the requirement is to detect a 1.0-dB loss anomaly with a confidence parameter of $a = 3$ (blue triangle), then the $\mathrm{SNR_{pp}}$ should be $> 23$ dB. Conversely, if an achievable $\mathrm{SNR_{pp}}$ is given, then the detectable limit of a loss is expressed by the inverse function of (32)

$$\mathrm{loss_{limit}}(z_m) = 1 - \frac{a}{\sqrt{\mathrm{SNR_{pp}}(z_m)}} \qquad (33)$$

### B. Required Sample Size and Optical Power

Since the expression for $\mathrm{SNR_{pp}}$ is obtained in the previous section, various quantitative designs of LPM is possible by substituting it into (32). Here, we provide examples of such designs, including the selection of sample size or optical power required to detect a given lumped loss under specific link and estimation conditions. Fig. 10 shows the required sample size to detect a given lumped loss for various optical power at arbitrary positions. To translate the required SNR to required sample, we used (22), where $\tilde{h}_r(\kappa)$ is obtained from the SCF for rectangular spectral signals. $a = 3$, $\beta_2 = -21.0$ ps²/km, $\gamma = 1.3$ W⁻¹km⁻¹, received SNR = 17.0 dB, $\Delta z = 1.0$ km, BW = 128 GHz are assumed. When the optical power decreases by 3 dB (corresponding to ~15 km for 0.2 dB/km), the number of samples required to achieve the same detection capability increases by a factor of four. Assuming that optimal launch power into fibers that maximizes a communication SNR may range from 0 to 3 dBm/ch for 128 GBd, the required sample size to detect a 1.0-dB loss anomaly at the beginning of a span is 2.1e5~8.5e5. To achieve the same accuracy at the end of a span with a 50-km fiber, 2.1e7~8.5e7 samples will be required.

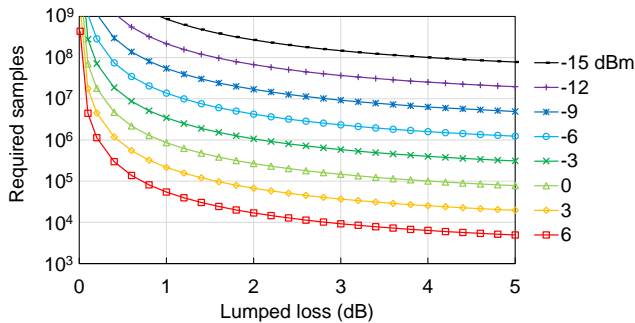

Fig. 10. Required sample size to detect a given lumped loss for various optical powers per channel. $a = 3$, $\beta_2 = $ -21.0 ps²/km, $\gamma = 1.3$ W⁻¹km⁻¹, received SNR = 17.0 dB, $\Delta z$ =1.0 km, BW = 128 GHz, and a rectangular spectral signal are assumed.

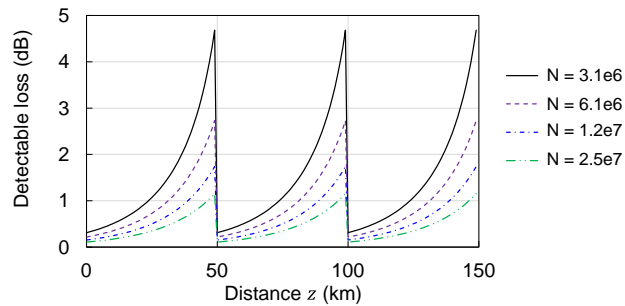

Fig. 11. Detectable lumped loss at arbitrary positions for various sample sizes. The conditions are the same as Fig. 5 and Fig. 6. $a = 3$, $\beta_2 = $-21.0 ps²/km, $\gamma = 1.3$ W⁻¹km⁻¹, $\Delta z = 1.0$ km, BW = 128 GHz, $L = 150$ km, launch power $P(0) = 2.0$ dBm, received SNR = 17.0 dB, and rectangular spectral signals are used.

Similarly, the required optical power can also be determined. If the sample size available is 1e7, the optical power should be around -5 dBm at a measurement position to detect a 1.0-dB loss. This implies that, to detect the loss at any positions in 50-km spans, a fiber launch power should be greater than 5 dBm. Note that these values are just examples and can also vary with other parameters such as $\Delta z$, $\beta_2$, $\gamma$, BW and $a$, according to the derived SNR (22). In this way, design parameters including sample size and optical power can be instantly selected without conducting massive experimental or numerical evaluations for possible link conditions.

### C. Dynamic Range

In the same way, it is possible to define the dynamic range (i.e., the tolerable span loss) of LPM for a given accuracy and estimation conditions. Let us assume that the sample size, fiber launch power, and detection capability should be 1e7, 2 dBm for 128GBd, and 2.0-dB loss with $a = 3$, respectively. Then from Fig. 10, -8dBm at the end of spans is required for successful detection everywhere in spans, implying the dynamic range is 10 dB.

### D. Position-wise Detectable Limit

Using (33), the detectable limit of a loss event at arbitrary positions can be specified. Fig. 11 is an example of the detectable limit under the same conditions as those shown in Fig. 5 and Fig. 6 with only the sample size varied based on (22). In this case, the launch power is set to 2 dBm/ch for 128 GBd. With a sample size of 6.1e6, a loss of 0.4 dB can be detected at the span input, while the capability is degraded to 2.7 dB at the span output. If 2.5e7 samples are available, then around 1.0-dB loss can be detected at any positions in 50-km spans. In this way, the accuracy of LPM at arbitrary positions can be quantitatively specified for various conditions. These results help to define the specifications of the LPM.

## IV. UNACCOUNTED EFFECTS

### A. Static distortion

In this paper, a lumped AWGN at the receiver was assumed, ignoring the impact of cross-channel interference (XCI), interaction of ASE and nonlinearity, and static distortions including transceiver imperfections or filtering penalties. While stochastic impairments such as the XCI and ASE-nonlinearity interaction can partly be considered embedded in the noise term $v$ and mitigated by the averaging effect, some static distortions may not vanish even with averaging or increasing the sample size. This means that there is a limit in improving the power-profile SNR, which is not accounted for in (22). To embed this SNR limit, one may need to consider decomposing the noise term $v$ into a stochastic term and a residual fixed term that cannot be reduced by the averaging effect.

### B. Modulation format

Two assumptions were made to obtain the expressions for the variance (21) and SNR (22): uniform variances over positions and stationary Gaussian signals. Based on these assumptions, the position-wise variance was represented by the trace metric and Re$[H^\dagger H]$ was considered a Toeplitz matrix, both of which promoted the analysis. However, the use of other modulation formats such as QPSK and 16QAM leads to the non-uniform variance and more noisy power profiles at the transmitter side. Under these practical formats, Re$[H^\dagger H]$ is not a mere Toeplitz but the magnitudes of the SCF in the first several rows are suppressed, as shown in Fig. 12. This figure is the QPSK counterpart of Fig. 3, where the rows were shown for a Gaussian signal format. This suppression effect was also reported in [32]. Due to this effect, the stronger deconvolution effect is brought by the inverse Re$[H^\dagger H]^{-1}$, particularly in the first several spans, resulting in greater noise enhancement in those areas. Consequently, the position-wise variance becomes non-uniform and increases near the transmitter side, as shown in Fig. 13, where the diagonals of Re$[H^\dagger H]^{-1}$ for various modulation formats are shown to represent the position-wise variance. In QPSK, the variance is nearly double that of Gaussian formats at the transmitter side, implying that twice the sample size is required to achieve the same accuracy as in the Gaussian case. Quantifying this increased variance under various formats and wider conditions requires further research.

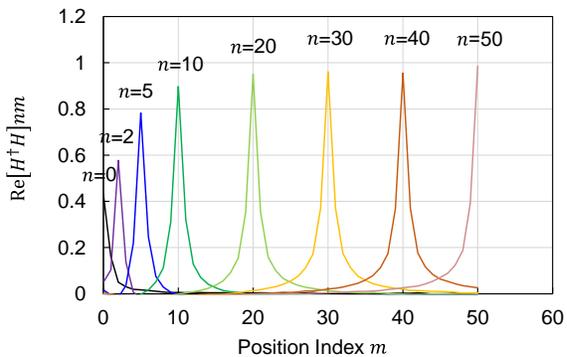

Fig. 12. $n$-th rows of $\text{Re}[H^\dagger H]$ for QPSK signal. $\beta_2$=-21.0 ps$^2$/km, BW = 128 GHz, $\Delta z$ = 1.0 km, $L$ = 50 km, and rectangular spectral signals are used.

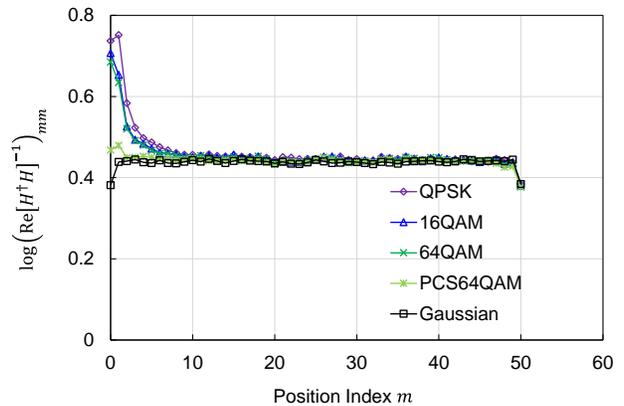

Fig. 13. Diagonal entries of $\text{Re}[H^\dagger H]^{-1}$ for various modulation formats. $\beta_2$=-21.0 ps$^2$/km, BW = 128 GHz, $\Delta z$ = 1.0 km, $L$ = 50 km, and rectangular spectral signals are used.

## C. Regularization

While we started with the more general estimator (8), our analysis mainly focused on a special case of $\lambda = 0$ for simplicity. Although we need to return to (11) to obtain the variance for $\lambda \neq 0$, the same logic applies as long as a regularization matrix $R$ is a Hermite Toeplitz or a closely related matrix and the resulting variance is uniform (e.g., $R = I$).

## V. Conclusion

This paper presented analytical results on the performance of LPM in the presence of noise and examples of the design of LPM. To formulate the accuracy of LPM at arbitrary positions along a link under given link and estimation conditions, the metrics such as the position-wise variance and SNR were first defined and analyzed. A notable thing is that these metrices can effectively be explained by the DTFT of the SCF, leading to an upper bound of the variance of LPM for a special case. It was also shown that the metric of the position-wise SNR defined in this paper directly explained the detectable limit of a loss event at individual positions. Based on these findings, examples of the design of LPM were shown, such as the selection of sample size and optical power required to achieve the requirement of the estimation accuracy. The performances such as the dynamic range and the detectable limit of loss events under given estimation conditions were also specified, facilitating the specifications of LPM.

As partly discussed in Section IV, there are several aspects of this analysis that require further research.

1. A lumped AWGN at the receiver is assumed, ignoring the impact of cross-channel interference, interaction of ASE and nonlinearity, and static distortions including transceiver imperfections or filtering penalties.
2. A stationary Gaussian signal is assumed, but other modulation formats such as QPSK, 16QAM, and PCS-QAM may necessitate modifications to the analysis presented here.
3. An upper bound of the variance is derived based on a Gaussian spectral signal: the extension for Nyquist signals is awaited.
4. The analytical results have only been shown to match simulations and need to be verified by experiment.

Nevertheless, the analysis presented here already captures essential behaviors of LPM in the presence of noise for a wide range of conditions and appear to be useful to design parameters and define the performance specification of LPM, as demonstrated in this paper.

## Appendix A

### Derivation of (11)

We first consider the covariance of $\text{Re}[G^\dagger \nu]$ and then derive that of $\text{Re}[G^\dagger G + \lambda R]^{-1}\text{Re}[G^\dagger \nu]$. Let $Z \sim \mathcal{CN}(\mu_Z, \Sigma_Z, C_Z)$ be a complex Gaussian vector, where $\mu_Z, \Sigma_Z, C_Z$ is a mean, covariance, and relation matrix of $z$. It is known that a linear transformation of $Z$ remains complex Gaussian:

$$AZ + b \sim \mathcal{CN}(A\mu_Z + b, A\Sigma_Z A^\dagger, AC_Z A^T) \quad (34)$$

In our case, $A = G^\dagger$, $Z = \nu$, and $b = 0$, and the noise $\nu$ in signals is assumed to be circularly symmetric, i.e., $\mu_\nu = 0$ and $C_\nu = 0$, yielding

$$G^\dagger \nu \sim \mathcal{CN}(0, G^\dagger \Sigma_\nu G) \quad (35)$$

Since the covariances of real-valued Gaussian vectors $X = \text{Re}[Z]$ satisfy $\Sigma_X = \frac{1}{2}\text{Re}[\Sigma_Z + C_Z]$ [28], we obtain

$$\text{Re}[G^\dagger \nu] \sim \mathcal{N}\left(0, \frac{1}{2}\text{Re}[G^\dagger \Sigma_\nu G]\right) \quad (36)$$

Applying a real-valued case of (34) for $A = \text{Re}[G^\dagger G + \lambda R]^{-1}$ gives (11).

## Appendix B

### Reference System Configuration

Throughout this paper, system parameters are frequently swept to validate analytical results. This appendix describes a reference system configuration and ranges of the parameter sweep. Two types of signals are used: Nyquist (rectangular) and Gaussian-spectral signals. The signal bandwidth (equivalent to symbol rate) is set to BW = 128 GHz for Nyquist signals, while $\sigma_\omega/2\pi$ = 54.4 GHz is used for Gaussian signals. The link under test consists of 3 spans of 50 km. To emulate fiber propagation, the split-step Fourier method was used with a spatial step size of 100 m with $\alpha = 0.20$ dB/km, $\beta_2 = -21.0$ ps$^2$/km and $\gamma = 1.30$ W$^{-1}$km$^{-1}$. A single polarization transmission was assumed. Lump noise with an SNR of 17 dB was added at the receiver. Upon reception, the linear solution $A_0(L)$ constructed using (4) was subtracted from the received signal to create $A_1(L)$ to perform the estimator (8). For the calculation of $G$, (5) and (6) are used.

Based on the reference configuration above, comprehensive simulations were conducted with possible combinations of the following parameters:

- CD coefficients $\beta_2 \in \{-5, -10, \dots -30\}$ ps$^2$/km ($\beta_2(z) = const., \beta_3(z) = 0$ were assumed)
- Total distance $L \in \{100, 250, 500\}$ km
- Signal bandwidth BW $\in \{64, 128, 256\}$ GHz for Nyquist signals, and $\sigma_w/2\pi \in \{27.2, 54.4, 109\}$ GHz for Gaussian spectral signals
- Spatial granularity $\Delta z \in \{0.5, 1, 2, 4\}$ km
- Signal samples N $\in [6.9e4, 5.0e6]$

Any deviations from these conditions are described in figure captions.

## Appendix C

### Derivation of (24) and (25)

(24) is the Fourier transform of the continuous SCF $h(z) = \frac{1}{\sqrt{1 + 2j(z/z_{CD}) + 3(z/z_{CD})^2}} = \frac{\sqrt{3}}{2} \frac{1}{\sqrt{1 + \left\{\frac{3}{2}\left(\frac{z}{z_{CD}} + \frac{1}{3}j\right)\right\}^2}}$ ( $z \in \mathbb{R}$ ). The Fourier transform of the square root of the Lorentzian is $\mathcal{F}\left[\frac{1}{\sqrt{1+t^2}}\right] = 2K_0(|\omega|)$, where $K_0(x)$ is the modified Bessel function of the second kind [36]. Applying a property of the Fourier transform $\mathcal{F}\left[f\left(\frac{t-t_0}{a}\right)\right] = ae^{-j\omega t_0}\tilde{f}(a\omega)$ for $h(z)$ gives (24).

(25) is the Fourier transform of the real part of the continuous SCF Re$[h(z)]$. Let us denote the SCF as follows

$$h(z) = \text{Re}[h(z)] + j\text{Im}[h(z)] \quad (37)$$

where Im$[\cdot]$ is the imaginary part. By applying the Fourier transform on both sides and due to its linearity,

$$\breve{h}(k) = \breve{h}_r(k) + j\breve{h}_i(k) \quad (38)$$

where $\breve{h}_r(k) = \mathcal{F}[\text{Re}[h(z)]]$ and $\breve{h}_i(k) = \mathcal{F}[\text{Im}[h(z)]]$. The complex conjugate of (37) is $h^*(z) = \text{Re}[h(z)] - j\text{Im}[h(z)]$, and its Fourier transform is

$$\breve{h}^*(-k) = \breve{h}_r(k) - j\breve{h}_i(k) \quad (39)$$

where $\mathcal{F}[h^*(z)] = \breve{h}^*(-k)$ is used. By adding (38) and (39), $\breve{h}_i(k)$ vanishes, yielding

$$\breve{h}_r(k) = \frac{\breve{h}(k) + \breve{h}^*(-k)}{2} \quad (40)$$

Substituting (24) into (40) yields (25).

## Appendix D

### Derivation of (28)

By substituting (27) into the integral along with $k = \kappa/\Delta z$,

$$\frac{1}{\pi}\int_0^\pi \frac{1}{\breve{h}_r(\kappa)}d\kappa \sim \sqrt{\frac{12}{\pi^3}}\frac{\Delta z}{z_{CD}}\int_0^\pi \sqrt{\frac{1}{3}\frac{z_{CD}}{\Delta z}}\kappa \exp\left(\frac{1}{3}\frac{z_{CD}}{\Delta z}\kappa\right)d\kappa \quad (41)$$

By substituting $u = \frac{1}{3}\frac{z_{CD}}{\Delta z}\kappa$, the right-hand side of (41) becomes

$$2\left(\frac{3}{\pi}\right)^{\frac{3}{2}}\frac{\Delta z^2}{z_{CD}^2}\int_0^{\frac{\pi z_{CD}}{3\Delta z}}\sqrt{u}\exp(u)\,du \quad (42)$$

Since $\int_0^c \sqrt{u}e^u du = \left|\Gamma\left(\frac{3}{2}, -c\right)\right|$, where $\Gamma(a, x)$ is the lower incomplete gamma function, we obtain (28).